\documentclass[preprin,showpacs,preprintnumbers,eqsecnum,amsmath,amssymb,nofootinbib]{revtex4}
\usepackage{graphics,epsfig,xcolor}
\usepackage{graphicx}
\usepackage{dcolumn}
\usepackage{bm}

\begin{document}
\baselineskip=0.8 cm
\title{Phantom hairy black holes and wormholes in Einstein-bumblebee gravity}

\author{Chikun Ding$^{1,2}$}\thanks{Corresponding author, email: dingchikun@163.com}
\author{Changqing Liu$^{1}$}\author{Yuehua Xiao$^{1}$}
\author{Jun Chen$^{1}$}
\affiliation{$^1$Department of Physics, Huaihua University, Huaihua, 418008, P. R. China
\\$^2$Key Laboratory of Low Dimensional Quantum Structures and Quantum Control of Ministry of Education,
 Hunan Normal University, Changsha, Hunan 410081, P. R. China
}

\vspace*{0.2cm}
\begin{abstract}
\baselineskip=0.6 cm
\begin{center}
{\bf Abstract}
\end{center}

In this paper we study Einstein-bumblebee gravity theory minimally coupled with external matter---a phantom/non-phantom(conventional) scalar field, and derive a series of hairy solutions---bumblebee-phantom(BP) and BP-dS/AdS black hole solutions, regular Ellis-bumblebee-phantom (EBP) and BP-AdS wormholes, etc. We first find that the Lorentz violation (LV) effect can change the so called black hole no-hair theorem and these scalar fields can  give a hair to a black hole. If LV coupling constant $\ell>-1$, the phantom field is admissible and the conventional scalar field is forbidden; if $\ell<-1$, the phantom field is forbidden and the conventional scalar field is admissible. By defining the Killing potential $\omega^{ab}$, we study the Smarr formula and the first law for the BP black hole, find that the appearance of LV can improve the structure of these phantom hairy black holes---the conventional Smarr formula and the first law of black hole thermodynamics still hold; but for no LV case, i.e., the regular phantom black hole reported in [Phys. Rev. Lett. {\bf96}, 251101], the first law cannot be constructed at all. When  the bumblebee potential is linear, we find that the phantom potential and the Lagrange-multiplier $\lambda$ behave as a cosmological constant $\Lambda$.

\end{abstract}

 \maketitle

\vspace*{0.2cm}
\section{Introduction}
The no-hair theorems \cite{herdeiro2015,robinson,chrusciel} state that in (electro-)vacuum general relativity, equilibrium black holes are very special celestial bodies which have only three unique quantities---mass, charge and angular momentum. These no-hair theorems are based on the assumptions of asymptotic flatness, the null energy condition \cite{nucamendi} and the existence of Gauss law \cite{herdeiro2015}.
However a scalar field with a non-trivial profile surrounding a black hole may modify the metric, lead to this black hole has a hair \cite{herdeiro}.
Scalar fields are the simplest matter which have been used to model dark energy and dark matter in cosmology and, may lead to "fifth" forces which can be constrained in a number of ways \cite{adelberger}. \textcolor[rgb]{0.00,0.00,1.00}{In 2021, Dimakis {\it et al} \cite{dimakis} derived an exact hairy black hole solution in Einstein-aether theory with a scalar field.} Other black holes with scalar hair have been constructed \cite{rao,benkenstein,feng,fan,herdeiroe2015,fan2016,lu}, and reviewed in \cite{herdeiro2015}.
Wormhole is another fascinating spacetime geometry which requires exotic matter and violates the null energy condition \cite{morris}. The simplest such exotic matter is the phantom scalar field, which is used to constructed the first traversable wormhole by Ellis in 1973 \cite{ellis}. Another phantom supported wormhole solutions have reported in Refs. \cite{gao2024,chew1,chew2,chew3,chew4}.

A phantom scalar field has a reversed sign of the kinetic energy and violates the null energy condition. To avoid the obvious quantum instability, it may be regarded as an effective field following from an underlying theory with positive energies \cite{nojiri}. In 2006, Bronnikov and Fabris studied a phantom/non-phantom scalar field with an arbitrary potential minimally coupling to the general relativity (GR) \cite{bronnikov}. They found that for the non-phantom(conventional) scalar field coupling, one cannot obtain wormholes or configurations ending with a regular 3-cylinder of finite radius (no-hair theorem). Then under a phantom scalar field coupling, they recovered the so-called Ellis wormhole and, derived a regular and asymptotically flat black hole solution. In 2013, we studied the strong gravitational lens effect of this regular black hole \cite{ding2013}. In 2023, Chataignier {\it et al} \cite{chataignier} studied an universes without singularities supported by this phantom field.  In 2019, \"{O}vg\"{u}n {\it et al} studied a wormhole solution with the isotropic matter in a Lorentz symmetry violation gravity theory---bumblebee gravity \cite{ovgun}. They found that under the condition of Lorentz symmetry violation, there exist a traversable wormhole solution with the conventional external matter.

In bumblebee gravity theory, the Lagrange density still preserves Lorentz invariance, but then the local Lorentz symmetry is just spontaneously violated by a potential $V(B_\mu B^\mu\mp b^2)$, where $B^\mu$ is a dynamic vector field (called the bumblebee field) with a nonzero vacuum expectation $b^\mu$ (a constant vector, $b^2=b_\mu b^\mu=$ constant), showing that the spacetime is anisotropic and has preferred frames\footnote{Note that another similar Lorentz symmetry violation theory is Einstein-aether theory, in which the Lorentz symmetry is spontaneously broken by an unitary and timelike everywhere vector $u^a$ (named the aether field, $u^a u_a=-1$) \cite{ding2015,ding2019,ding2021,wang2020}. One can also introduces a dynamic tensor field $B^{\mu\nu}$ (called the  Kalb-Ramond field), then the spontaneous Lorentz violation is triggered by a potential $V(B_{\mu\nu} B^{\mu\nu}\mp b^2)$ for an antisymmetric 2-tensor field $B^{\mu\nu}$ with a nonzero vacuum expectation $b^{\mu\nu}$ (a constant tensor field, $b^2=b_{\mu\nu}b^{\mu\nu}$= constant) \cite{yang,brett}.}. The underlying geometry is assumed to Riemannian type(purely metric), the non-Riemannian treatment (metric-affine) had been considered in Ref. \cite{delhom,filho}. This bumblebe field looks like the electromagnetical vector potential $A^\mu$, but here $A^\mu$ has a non-zero constant vacuum expectation $a^\mu$ \footnote{The authors' in Ref. \cite{xu} concluded that the bumblebee model can recover the Einstein-Maxwell theory when bumblebee coupling constant $\xi$=0, i.e., the Reissner-Nordst\"{o}m solution. It seems incorrect because that in their studies, the norm $b^\mu b_\mu$ is not a constant.}. The early study \cite{kostelecky198940} on this model was just using Einstein-Maxwell system with a potential $V(A_\mu A^\mu-a^2)$, where $a^2=a_\mu a^\mu$ was a positive  constant.

Studying the Lorentz symmetry breaking is a useful approach toward investigating the foundations of modern physics. There are many theoretical models concern Lorentz invariance violation (LV), such as the standard model extension \cite{kostelecky2004}, string theory \cite{kostelecky198939}, noncommutative field theories \cite{carroll,carroll2,carroll3}, massive gravity \cite{fernando}, etc. The surprising property of this bumblebee gravity is that it does not forbid the propagation of massless vector modes \cite{bluhm2008}. Therefore, one expects to reveal a variety of physical relics used in studies of dark energy and dark matter due to the appearance of Nambu-Goldstone and massive Higgs in this LV theory \cite{Kostelecky6,Kostelecky7,bluhm}.

We will study the possible wormhole solution and black hole solution in this Einstein-bumblebee gravity theory with an external matter---phantom/non-phantom field, try to find the effect of the Lorentz invariance violation on the conventional black hole physics.  The rest paper is organized as follows.   In Sec. II we give the background for the Einstein-bumblebee theory and derive the gravitational field equations, bumblebee motion equations and scalar motion equations. In Sec. III, we derive some wormhole and black hole solutions, give the form of phantom charge $Q_p$ and corresponding potential $V_p$.  In Sec. IV, we construct the Smarr formula and the first law of black hole thermodynamics for the bumblebee phantom (BP) black hole and find the effect of Lorentz violation on black hole thermodynamics. Sec. V for some astrophysical observations of BP black hole. Sec. VI for a  summary.

\section{Einstein-bumblebee gravity minimally coupled with phantom scalar field}

In the bumblebee gravity model, one introduces the bumblebee dynamical vector field $B_{\mu}$ which has a nonzero vacuum expectation value, leading to a spontaneous Lorentz symmetry breaking in the gravitational sector via a given potential. The action is \cite{ding2021},
\begin{eqnarray}
\mathcal{S}=
\int d^4x\sqrt{-g}\Big[\frac{1}{2\kappa}\big(R+\varrho B^{\mu}B^{\nu}R_{\mu\nu}\big) -\frac{1}{4}B^{\mu\nu}B_{\mu\nu}
-V(B_\mu B^{\mu}\mp b^2)+\mathcal{L}_M\Big], \label{action}
\end{eqnarray}
where $R$ is the Ricci scalar \footnote{Note that in the Ref. \cite{dzhunushaliev}, there is a minus sign before the Ricci scalar, i.e., $-R$. This is because that the Ricci tensor is defined by the contraction of the first and third index of the Riemann tensor $R_{\mu\nu}=R^\sigma_{\;\mu\sigma\nu}$ there. However, here we apply the contraction of the first and fourth index, $R_{\mu\nu}=R^\sigma_{\;\mu\nu\sigma}$, so there is no that minus sign, see the appendix for detail.}, $\mathcal{L}_M$ is the external matter field,  $\kappa=8\pi G/c^4$, where $G$ is the four dimensional Newton's gravitational constant. Here and hereafter, we take $G=1$ and $c=1$ for convenience.

The coupling constant $\varrho$ dominates the non-minimal gravity interaction to bumblebee field $B_\mu$ and \begin{eqnarray}\varrho\neq0 \text{ and } \varrho\neq-1/b_0^2,\end{eqnarray} where  $b_0$ is a positive constant relative to the following constant vector $b^\mu$.
The potential $V(B_\mu B^{\mu}\mp b^2)$ triggers Lorentz and/or $CPT$ (charge, parity and time)
violation, where the dynamical field $B_{\mu}$ acquires a nonzero vacuum expectation value (VEV), $\langle B^{\mu}\rangle= b^{\mu}$,
satisfying the condition $B^{\mu}B_{\mu}\pm b^2=0$, where $b^2=b_{\mu}b^{\mu}$ is the norm of the  constant vector $b^\mu$.
 The constant vector $b^{\mu}$ is a function of the spacetime coordinates and has a constant value
 $b_{\mu}b^{\mu}=\mp b_0^2$, where $\pm$ signs mean that $b^{\mu}$ is timelike or spacelike,
 respectively. It gives a nonzero  VEV for bumblebee field $B_{\mu}$
 indicating that the vacuum of this model obtains a prior direction in the spacetime.
The bumblebee field strength is
\begin{eqnarray}
B_{\mu\nu}=\partial_{\mu}B_{\nu}-\partial_{\nu}B_{\mu}.
\end{eqnarray}
This antisymmetry of $B_{\mu\nu}$ implies the constraint \cite{bluhm}
\begin{eqnarray}
\nabla ^\mu\nabla^\nu B_{\mu\nu}=0.
\end{eqnarray}

Black hole no-hair theorem states that stationary black  hole solutions are hairless except that three hairs: mass, angular momentum and charge. In this paper, we consider a counter example---
minimally coupling this bumblebee gravity theory with an external matter field,
\begin{eqnarray}
\mathcal{L}_M=\frac{1}{2\kappa}\big[\varepsilon\partial_\mu\Phi\partial^\mu\Phi
-2\mathcal{V}(\Phi)\big], \label{action}
\end{eqnarray}
where $\Phi$ is a scalar field, $\mathcal{V}$ is its scalar  potential, $\varepsilon=+1$ for a usual scalar field with positive kinetic energy, $\varepsilon=-1$ for a phantom field with negative kinetic energy.

Varying the action (\ref{action}) with respect to the metric yields the gravitational field equations
\begin{eqnarray}\label{einstein0}
G_{\mu\nu}=\kappa T_{\mu\nu}^B+T_{\mu\nu}^p,
\end{eqnarray}
where the Einstein's tensor $G_{\mu\nu}=R_{\mu\nu}-g_{\mu\nu}R/2$,
$T_{\mu\nu}^p$ is the energy momentum tensor of the scalar field,
\begin{eqnarray}
T_{\mu\nu}^p=-\varepsilon\partial_\mu\Phi\partial_\nu\Phi+\frac{\varepsilon}{2}g_{\mu\nu}\partial_\rho\Phi\partial^\rho
\Phi-g_{\mu\nu}\mathcal{V}(\Phi).
\end{eqnarray}
The bumblebee energy momentum tensor $T_{\mu\nu}^B$ is
\begin{eqnarray}\label{momentum}
&&T_{\mu\nu}^B=B_{\mu\alpha}B^{\alpha}_{\;\nu}-\frac{1}{4}g_{\mu\nu} B^{\alpha\beta}B_{\alpha\beta}- g_{\mu\nu}V+
2B_{\mu}B_{\nu}V'\nonumber\\
&&+\frac{\varrho}{\kappa}\Big[\frac{1}{2}g_{\mu\nu}B^{\alpha}B^{\beta}R_{\alpha\beta}
-B_{\mu}B^{\alpha}R_{\alpha\nu}-B_{\nu}B^{\alpha}R_{\alpha\mu}\nonumber\\
&&+\frac{1}{2}\nabla_{\alpha}\nabla_{\mu}(B^{\alpha}B_{\nu})
+\frac{1}{2}\nabla_{\alpha}\nabla_{\nu}(B^{\alpha}B_{\mu})
-\frac{1}{2}\nabla^2(B_{\mu}B_{\nu})-\frac{1}{2}
g_{\mu\nu}\nabla_{\alpha}\nabla_{\beta}(B^{\alpha}B^{\beta})\Big],
\end{eqnarray}
where the prime denotes differentiation with respect to the argument,
\begin{eqnarray}
V'=\frac{\partial V(x)}{\partial x}\Big|_{x=B^{\mu}B_{\mu}\pm b^2}.
\end{eqnarray}
Varying instead with respect to the bumblebee field generates the bumblebee equations of motion (supposing that there is no coupling between the bumblebee field and the phantom field),
\begin{eqnarray}\label{motion}
\nabla ^{\mu}B_{\mu\nu}=2V'B_\nu-\frac{\varrho}{\kappa}B^{\mu}R_{\mu\nu}.
\end{eqnarray}
Varying with respect to the scalar field generates the  scalar equation of motion,
\begin{eqnarray}\label{motion3}
\varepsilon\frac{1}{\sqrt{-g}}\partial_\mu\big(\sqrt{-g}g^{\mu\nu}\partial_\nu\Phi\big)+\frac{d\mathcal{V}}{d\Phi}=0.
\end{eqnarray}

The contracted Bianchi identities ($\nabla ^\mu G_{\mu\nu}=0$) lead to conservation of the total energy-momentum tensor
\begin{eqnarray}\label{}
\nabla ^\mu T_{\mu\nu}=\nabla ^\mu\big( T^B_{\mu\nu}+T^p_{\mu\nu})=0.
\end{eqnarray}
We suppose that the bumblebee field is frozen at its VEV like in Refs \cite{casana,bertolami}, i.e., it is
\begin{eqnarray}\label{bbu}
B_\mu=b_\mu.
\end{eqnarray}
And the potential has a smooth quadratic function, \begin{eqnarray}
V=\frac{k}{2}x^2, \qquad x=(B_\mu B^\mu-b^2),
\end{eqnarray}
where $k$ is a constants and that minimized by the condition $x=0$; or a linear Lagrange-multiplier function form \cite{bluhm}
\begin{eqnarray}\label{linep}
V=\frac{\lambda}{2}x,
\end{eqnarray}
 where $\lambda$ is a non-zero constant and deserves as a Lagrange-multiplier field which is auxiliary and has no kinetic terms.
The linear potential (\ref{linep}) is $V=0$ under the condition (\ref{bbu}) and its derivative is $V'=\lambda/2$ which can modify the Einstein equations. However, this additional degree of freedom of the $\lambda$ field is auxiliary. Then the first two terms in Eq. (\ref{momentum}) are like those of the electromagnetic field, the only distinctiveness are the coupling items to Ricci tensor and this $\lambda$ term. Under this condition,  Eq. (\ref{einstein0}) leads to gravitational field equations \cite{ding2021}
\begin{eqnarray}\label{bar}
G_{\mu\nu}=C_{\mu\nu}+\bar B_{\mu\nu}+T^p_{\mu\nu},
\end{eqnarray}
with
\begin{eqnarray}\label{barb}
&&C_{\mu\nu}=\kappa (2V' b_\mu b_\nu+b_{\mu\alpha}b^{\alpha}_{\;\nu}-\frac{1}{4}g_{\mu\nu} b^{\alpha\beta}b_{\alpha\beta})+\varrho\Big(\frac{1}{2}
g_{\mu\nu}b^{\alpha}b^{\beta}R_{\alpha\beta}- b_{\mu}b^{\alpha}R_{\alpha\nu}
-b_{\nu}b^{\alpha}R_{\alpha\mu}\Big),\\
&&\bar B_{\mu\nu}=\frac{\varrho}{2}\Big[
\nabla_{\alpha}\nabla_{\mu}(b^{\alpha}b_{\nu})
+\nabla_{\alpha}\nabla_{\nu}(b^{\alpha}b_{\mu})
-\nabla^2(b_{\mu}b_{\nu})-g_{\mu\nu}\nabla_\alpha\nabla_\beta(b^\alpha b^\beta)\Big].
\end{eqnarray}

The static spherically symmetric black hole metric has the form\begin{eqnarray}\label{metric}
&&ds^2=e^{2\phi(\rho)}dt^2-e^{2\psi(\rho)}d\rho^2-r^2(\rho)d\Omega^2,
\end{eqnarray}
where $\Omega$ is a standard two-sphere $d\Omega^2=d\theta^2+\sin^2\theta d\varphi^2$ and the scalar field $\Phi=\Phi(\rho)$.
In this static spherically symmetric spacetime, the most general form for the bumblebee field would be $b_\mu=(b_t,b_\rho,0,0)$, where $b_t$ and $b_\rho$ are functions of $\rho$ subject to the constraint $-b_t^2 e^{ -2\phi}+ b_\rho^2 e^{-2 \psi} = b_0^2$, here $b_0$ is a positive constant. In this general case, the bumblebee field has both radial and a time component for the vacuum expectation value. In the purely radial case $b_t=0$, the authors in Refs. \cite{bertolami,casana} obtained new black hole solutions indeed. But for the general case(temporal and radial), the authors in Ref. \cite{bertolami} obtain a slightly perturbed metric, where one cannot constrain the physical parameters from the observed limits on the PPN(parameterized post-Newtonian) parameters. Hence here we consider only the purely radial case to get a black hole solution and let the general case for the future work.

We pay attention to that the bumblebee field has a radial vacuum energy expectation because that the spacetime curvature has a strong radial variation, on the contrary that the temporal changes are very slow. Now the bumblebee field is supposed to be spacelike  as that
\begin{eqnarray}\label{bu}
b_\mu=\big(0,b_0e^{\psi(\rho)},0,0\big).
\end{eqnarray}
 Then the bumblebee field strength is
\begin{eqnarray}
b_{\mu\nu}=\partial_{\mu}b_{\nu}-\partial_{\nu}b_{\mu},
\end{eqnarray}
whose components are all zero. And their divergences are all zero, i.e.,
\begin{eqnarray}
\nabla^{\mu}b_{\mu\nu}=0.
\end{eqnarray}
From the bumblebee field motion equation (\ref{motion}), we have the projection of the Ricci tensor along the bumblebee field is
\begin{eqnarray}
\varrho b^{\mu}R_{\mu\nu}=2\kappa V'b_\nu\label{motion2},
\end{eqnarray}
with the radial component of Ricci tensor $R_{11}$
\begin{eqnarray}\label{ricc}
R_{11}=-2\frac{r'}{r}\psi'+2\frac{r''}{r}+(\phi''-\phi'\psi'+\phi'^2)\label{R11}.
\end{eqnarray}
From Eq. (\ref{motion2}), one can see that the bumblebee coupling constant $\varrho$ can't be zero, i.e.,  $\varrho\neq0$: if $V'=0$, then it hasn't any restriction on $R_{\mu\nu}$ when $\varrho=0$, the system goes back to GR; if $V'\neq0$, then $\varrho=0$ compels $b_\nu=0$, the system goes also back to GR.

As to gravitational field equation (\ref{bar}), one can obtain the following three  component equations
\begin{eqnarray}
&&-\frac{e^{2\psi}}{r^2}+(1+\ell)\big(-\frac{2}{r}r'\psi'+\frac{r'^2}{r^2}+\frac{2}{r}r''\big) -\ell R_{11}=-\frac{\varepsilon}{2}\Phi'^2-e^{2\psi}\mathcal{V},\label{tt}\\
&&\frac{e^{2\psi}}{r^2}-(1+\ell)\big(\frac{2}{r}r'\phi'+\frac{r'^2}{r^2}\big) +\ell  \frac{2\kappa V'}{\varrho}e^{2\psi} =-\frac{\varepsilon}{2}\Phi'^2+e^{2\psi}\mathcal{V},\label{rho}\\
&&(1+\ell)\big[\frac{r''}{r}-\frac{r'}{r}(\phi'+\psi')\big] -R_{11}=\frac{\varepsilon}{2}\Phi'^2+e^{2\psi}\mathcal{V},\label{theta}
\end{eqnarray}
where we have redefined the Lorentz-violating parameter $\ell=\varrho b_0^2$ and, the prime $'$ is the derivative with respect to the corresponding argument, respectively.
Adding the Eq. (\ref{tt}) to (\ref{theta}), one can obtain that
\begin{eqnarray}\label{}
-\frac{1}{r^2}e^{2\psi}+(1+\ell)\left[\frac{r''}{r}+\frac{r'^2}{r^2}
-\frac{r'}{r}(\phi'+\psi')-(\phi''-\phi'\psi'+\phi'^2)\right]=0,\label{theta2}
\end{eqnarray}
which doesn't dependent on the scalar field but on the bumblebee coupling constant $\ell$. It is easy to see that if the bumblebee coupling constant $\ell=-1,  ($or$ \; \varrho=-1/b_0^2)$, then $e^{2\psi}=0$, which is nonphysical and unacceptable. Therefore, the coupling constant $\ell\neq0$ and $\ell\neq-1$ are assumed. For the static spacetime, one can set that $\phi'+\psi'=0$. Here for convenient, we let $e^{2\phi}=A(\rho)$ and $e^{2\psi}=(1+\ell)/A(\rho)$, so Eq. (\ref{theta2}) can be simplified as
\begin{eqnarray}\label{PRL96}
A(\rho)(r^2)''-r^2A''(\rho)=2,
\end{eqnarray}
which is the same as that Eq. (6) in Ref. \cite{bronnikov}. And the Ricci component $R_{11}$ (\ref{ricc}) becomes
\begin{eqnarray}\label{}
R_{11}=\frac{A''(\rho)}{2A(\rho)}+2\frac{r''}{r}+\frac{r'A'(\rho)}{rA(\rho)}.
\end{eqnarray}
One can solve the Eq. (\ref{PRL96}) with a given function $r(\rho)$ to obtain the black hole solution $A(\rho)$, for example, one can choose $r=\sqrt{\rho^2+p^2},p=$const$>0$, then the solution Eq. (10) in Ref.  \cite{bronnikov} can be recovered when there is no bumblebee field, i.e., the bumblebee coupling constant $\ell\rightarrow0$. However, there is another restricted equation---the bumblebee motion equation (\ref{motion2})  which needs to be considered,
\begin{eqnarray}\label{restrict}
\frac{A''(\rho)}{2A(\rho)}+2\frac{r''}{r}+\frac{r'A'(\rho)}{rA(\rho)}=-(1+\ell)\frac{2\kappa V'}{\varrho A(\rho)}.
\end{eqnarray}
Combination of Eq. (\ref{PRL96}) and (\ref{restrict}) gives that,
\begin{eqnarray}\label{Aequation}
\left[\frac{2\kappa V'}{\varrho}(1+\ell)-\frac{1}{r^2}\right]+\left(\frac{2r''}{r}+\frac{1}{r^2}\right)A+
\frac{r'}{r}A'=0.
\end{eqnarray}
Eqs. (\ref{tt}), (\ref{rho}) and (\ref{motion3})  can give the phantom field, its potential and phantom motion equation as those
\begin{eqnarray}\label{scalarp}
-\varepsilon\Phi'^2=2(1+\ell)\Big(\frac{r''}{r}+2b_0^2\frac{\kappa V'}{A}\Big),\qquad \mathcal{V}=-\frac{1}{2r^2}(r^2A')',\qquad \frac{\varepsilon(Ar^2\Phi')'}{1+\ell}=r^2\frac{d\mathcal{V}}{d\Phi},
\end{eqnarray}
which are the same as those Eqs. (3), (4) and (5) in Ref. \cite{bronnikov} when coupling constant $\ell=0$ or $b_0=0$.

From the above equations (\ref{Aequation}), and (\ref{scalarp}), one can conclude that some consequences about the no-hair theorems: (i), if $\ell>-1$ \footnote{\textcolor[rgb]{0.00,0.00,1.00}{For the hairy black hole solution in Einstein-aether theory \cite{dimakis}, if $\mu>0$, the external scalar field $\phi(r)$ is imaginary to be of a phantom field; if $-1/2<\mu<0$, the scalar field is conventional.}}, then the $\varepsilon=-1$ phantom field can exist but the $\varepsilon=+1$ scalar field with positive kinetic energy is forbidden by the given spacetime due to that $r''\geq0$; (ii), if bumblebee coupling constant $\ell<-1$ and $V'=0$ \footnote{In Ref. \cite{ovgun}, the authors considered that when the bumblebee coupling constant $\ell=-2$, the energy density $\tilde{\rho}$ of the  matter is conventional positive, (see the appendix for details).}, then the $\varepsilon=+1$ scalar field with positive kinetic energy can exist but $\varepsilon=-1$ phantom field is forbidden; (iii), the bumblebee field affects the spacetime via the coupling constant $\ell$ and the bumblebee motion equqtion (\ref{restrict}); (iv), from Eqs. (\ref{PRL96}) and (\ref{scalarp}), it seems that the scalar field can't affect the gravitational field equation and on the contrary, the contents of scalar field and the forms of its potential $\mathcal{V}$ are determined by the given spacetime. However, this scalar field can give a hair to a black hole via $r''(\rho)$ and $V'$ in Eq. (\ref{scalarp}). In the next section, we will consider the $\varepsilon=-1$ phantom field and the bumblebee coupling constant $\ell>0$, and derive some exact wormhole and black hole solutions in this Lorentz violating theory.

\section{Phantom wormhole and  hairy black hole solutions}

In this section, we will derive an Ellis like, Schwarzschild like wormhole, some asymptotic to AdS(anti-de-Sitter) wormholes, some  bumblebee-phantom(BP)-flat black holes and some  BP-dS/AdS black holes under the two kinds of bumblebee potential $V=kx^2/2$ and $V=\lambda x/2$.
\subsection{Case I:  quadratic bumblebee potential}Let $r^2=\rho^2+p^2, p=$const$>0$ and $V=kx^2/2$, then Eq. (\ref{Aequation}) becomes
\begin{eqnarray}\label{}
\rho A'(\rho)+\frac{\rho^2+3p^2}{r^2}A(\rho)-1=0.
\end{eqnarray}
One can obtain that,
\begin{eqnarray}\label{metricc}
A(\rho)=\big(1+\frac{p^2}{\rho^2}\big)\Big[1+\frac{\rho_0}{\rho}-\frac{p}{\rho}\arctan\frac{\rho}{p}\Big],
\end{eqnarray}
where $\rho_0$ is an integral constant.
The metric (\ref{metric}) becomes,
\begin{eqnarray}\label{metric1}
&&ds^2=A(\rho)dt^2-\frac{1+\ell}{A(\rho)}d\rho^2-(\rho^2+p^2)d\Omega^2.
\end{eqnarray}
The phantom field and its potential are,
\begin{eqnarray}\label{ppotental}
\Phi=\pm\sqrt{2(1+\ell)}\arctan\frac{\rho}{p}+\Phi_0,\qquad \mathcal{V}=\frac{2p^2}{(\rho^2+p^2)^2}A(\rho),
\end{eqnarray}
which are asymptotical to be as $\Phi\rightarrow$constant and $\mathcal{V}\rightarrow0$ at $\rho\rightarrow\pm\infty$.
The behaviors of metric function $A(\rho)$ listed in Tab. I.
 \begin{table}
 \caption{Asymptotic behaviors of metric function $A(\rho)$ in Einstein-bumblebee gravity coupled with a phantom field under the quadratic bumblebee potential $V=kx^2/2$.}\label{tab1}
\begin{center}
\begin{tabular}{|c|c|c|c|}
\hline  $\rho_0$ & large $\rho$ & little $\rho$ & kinds of solution   \\
\hline $\rho_0=0$ & $\rho\rightarrow\pm\infty, A(\rho)\rightarrow1$ & $\rho\rightarrow\pm0, A(\rho)\rightarrow1/3$& EBP wormhole\\
$\rho_0>0$ & $\rho\rightarrow\pm\infty, A(\rho)\rightarrow1$ & $\rho\rightarrow\pm0, A(\rho)\rightarrow\pm\infty$&Schwarzschild-like wormhole\\
$\rho_0<0$ & $\rho\rightarrow\pm\infty, A(\rho)\rightarrow1$ & $\rho\rightarrow\pm0, A(\rho)\rightarrow\mp\infty$& BP black hole\\
\hline
\end{tabular}
\end{center}
\end{table}
It shows that the metric function $A(\rho)$ of the solutions are all asymptotically to unit, i.e., $A(\rho)\rightarrow1$, and constant $\rho_0\geq0$ corresponding to wormhole solution,  $\rho_0<0$ for black hole solution. Choosing in (\ref{ppotental}) the plus sign and $\Phi_0=0$, we obtain for $\mathcal{V}$ ($\Psi:=\Phi/\sqrt{2(1+\ell)}$ ),
\begin{eqnarray}\label{}
\mathcal{V}=\frac{2}{p^2}\cos^2\Psi\cot^2\Psi\big[1+\cot\Psi(\frac{\rho_0}{p}-\Psi)\big],
\end{eqnarray}
which has at least two zero-slope points at different values of $\Phi$ as those in Ref. \cite{bronnikov}.

\subsubsection{Ellis-bumblebee-phantom(EBP) wormhole solution}
If $\rho_0=0$, then $A(\rho)$ becomes
\begin{eqnarray}\label{wormhole}
A(\rho)=\big(1+\frac{p^2}{\rho^2}\big)\Big(1-\frac{p}{\rho}
\arctan\frac{\rho}{p}\big).
\end{eqnarray}
In the little $\rho$ region, it can be expanded as,\begin{eqnarray}\label{}
A(\rho)=\frac{p^2}{\rho^2}\Big(\frac{\rho^2}{3p^2}
-\frac{\rho^4}{5p^4}+\cdots\big)=\frac{1}{3}+\mathcal{O}(\rho^2).
\end{eqnarray}
Therefore, when $\rho\rightarrow\pm0$, $A(\rho)$ is bounded by $1/3$. In the large $\rho$ region, it can be expanded as,
\begin{eqnarray}\label{}
A(\rho)=\big(1+\frac{p^2}{\rho^2}\big)\Big[1-\frac{\pi p}{2\rho}+\frac{p^2}{\rho^2}-\frac{1}{3}\frac{p^4}{\rho^4}
+\cdots\Big]=1-\mathcal{O}(\frac{1}{\rho}).
\end{eqnarray}
So  in  all region $-\infty<\rho<\infty,$ the metric function $A(\rho)$ is regular and positive everywhere, i.e., $ 1/3<A(\rho)<1$. Then it isn't a black hole solution due to that it has no horizon. It is a wormhole solution with throat radius $r=p, (\rho=0)$. This throat connects two asymptotically flat regions of spacetime. One can see that it is asymptotic to an Ellis wormhole \cite{ellis} \footnote{For Ellis wormhole, its metric is $ds^2=dt^2-d\rho^2-(\rho^2+p^2)d\Omega^2$. Using coordinate transformation $\rho=\pm\sqrt{r^2-p^2}$, it can be rewritten as $ds^2=dt^2-\frac{dr^2}{1-p^2/r^2}-r^2d\Omega^2$.}, and it can be called as Ellis-bumblebee-phantom (EBP) wormhole.
 Its Ricci scalar $R$ is
\begin{eqnarray}\label{}
R=\frac{2}{(1+\ell)(\rho^2+p^2)}\Big[\ell-\frac{p^2}{\rho^2}\big(1+\frac{6p^2}{\rho^2}\big)
+\frac{3p^3}{\rho^3}\big(1+\frac{2p^2}{\rho^2}\big)\arctan\frac{\rho}{p}
\Big].
\end{eqnarray}
When at origin $\rho\rightarrow0$, it becomes
\begin{eqnarray}\label{}
R=\frac{2(5\ell+6)}{5(1+\ell)p^2},
\end{eqnarray}
which is finite. Then this EPB wormhole has no singularity everywhere.

From the Eq. (\ref{wormhole}), when $p\rightarrow0$, then $A(\rho)=1, \Phi=$ constant, $\mathcal{V}=0$, the spacetime metric becomes $ds^2=dt^2-(1+\ell)d\rho^2-\rho^2d\Omega^2$ \footnote{Note that in Ref. \cite{ovgun}, if the bumblebee motion equation $R_{rr}=0$ is used, then this metric $ds^2=dt^2-(1+\ell)d\rho^2-\rho^2d\Omega^2$ can also be obtained.}.  It isn't a wormhole, and means that a wormhole needs an exotic matter to sustain. This metric has a nonzero Ricci scalar $R=2\ell/(1+\ell)\rho^2$, therefore, it isn't Minkowski spacetime.
\subsubsection{Schwarzwschild-like wormhole solution}
When $\rho_0=\pi p/2>0$, then $A(\rho)$ becomes,
\begin{eqnarray}\label{}
A(\rho)=\big(1+\frac{p^2}{\rho^2}\big)\Big[1+\frac{p}{\rho}\big(\frac{\pi}{2}
-\arctan\frac{\rho}{p}\big)\Big],
\end{eqnarray}
which is a Schwarzschild-like wormhole solution and the singularity $\rho=0$ becomes naked \footnote{It likes the negative mass Schwarzschild solution which has no horizon and the singularity is naked \cite{visser}.}.
\subsubsection{Bumblebee-phantom(BP) black hole solution}
If $\rho_0=\pi p/2-2M<0$, then $A(\rho)$ becomes,
\begin{eqnarray}\label{metricf}
A(\rho)=\big(1+\frac{p^2}{\rho^2}\big)\Big[1-\frac{2M}{\rho}+\frac{p}{\rho}\big(\frac{\pi}{2}
-\arctan\frac{\rho}{p}\big)\Big],
\end{eqnarray}
where $M$ is the ADM(Arnowitt-Deser-Misner) mass of the black hole (ADM mass is identical to the Komar mass defined with the time translation Killing vector $\xi^\mu$ \cite{ding2015}). From the metric (\ref{metric1}), one can see that when $p\rightarrow0, \ell\rightarrow0, A(\rho)=1-2M/\rho$, the Schwarzschild black hole solution is recovered; when $p\rightarrow0$, the bumblebee Schwarzschild-like black solution \cite{casana} is recovered and also, the phantom field $\Phi\rightarrow $constant and the potential $\mathcal{V}\rightarrow0$ \footnote{Note that in Ref. \cite{bronnikov}, the regular phantom black hole can also be back to Schwarzschild case. This solution is \cite{ding2013}, \begin{eqnarray}\label{}
A(\rho)=1-\frac{3M}{p}\Big[\big(1+\frac{\rho^2}{p^2}\big)\big(\frac{\pi}{2}
-\arctan\frac{\rho}{p}\big)-\frac{\rho}{p}\Big],\nonumber
\end{eqnarray}
 in which when $p\rightarrow0, A(\rho)\rightarrow 1-2M/\rho$. }. When $\rho$ is large, the metric function (\ref{metricf}) is asymptotical to the Schwarzschild black hole solution, $A(\rho)\rightarrow1-2M/\rho$ and, the phantom field $\Phi\rightarrow $constant and the potential $\mathcal{V}\rightarrow0$. This black hole can be named as bumblebee-phantom(BP) black hole. It has a single horizon,
\begin{eqnarray}\label{horizon}
\rho_h=2M-p\big(\frac{\pi}{2}-\arctan\frac{\rho_h}{p}\big),
\end{eqnarray}
and the Hawking temperature is,
\begin{eqnarray}\label{}
T=\frac{1}{4\pi\sqrt{1+\ell}\rho_h}.
\end{eqnarray}
By constructing a Komar integral \cite{ding2023}, one can obtain the Smarr formula (see the Sec. IV for detail),
\begin{eqnarray}\label{}
M=2TS+2Q_pV_p,\quad\text{with}\quad Q_p=\sqrt{2(1+\ell)}p,\;V_p=\frac{1}{2\sqrt{2(1+\ell)}}
\Big[\frac{\pi}{2}-\arctan \frac{\rho_h}{p}\Big],
\end{eqnarray}
and the first law,
\begin{eqnarray}\label{}
dM=TdS+V_pdQ_p,
\end{eqnarray}
where $S=\pi\sqrt{(1+\ell)}(\rho_h^2+p^2)$ is the BP black hole's entropy, $Q_p$ is the phantom charge, $V_p$ is the corresponding potential (an effective potential). It easy to see that this black hole does contain phantom scalar charge like those scalar hairy black holes in Ref. \cite{herdeiro,dimakis}. \textcolor[rgb]{0.00,0.00,1.00}{ The authors in Ref. \cite{dimakis} derived a hairy black hole solution with a conventional/pahantom scalar field coupled with a spacelike aether field $u^a$ in Einstein-aether theory which is also a Lorentz violating theory. In their solution, the Lorentz violating constant is $\mu$ which appears as a measure of the radial modification due to the aether having a velocity in the $r$ direction; the constant $C$ shows the content of the scalar field.}

\subsection{Case II: linear bumblebee potential}
Let $r^2=\rho^2+p^2, p=$const$>0$ and the bumblebee potential $V=\lambda x/2$ is linear, then Eq. (\ref{Aequation}) becomes
\begin{eqnarray}\label{}
\rho A'(\rho)+\frac{\rho^2+3p^2}{r^2}A(\rho)+(1+\ell)\frac{\kappa \lambda}{\varrho}r^2-1=0.
\end{eqnarray}
One can obtain that,
\begin{eqnarray}\label{}
A(\rho)=\big(1+\frac{p^2}{\rho^2}\big)\Big[1+\frac{\rho_1}{\rho}-\frac{ \Lambda}{3}\rho^2-\frac{p}{\rho}\arctan\frac{\rho}{p}\Big],
\end{eqnarray}
where $\rho_1$ is an integral constant and $\Lambda=(1+\ell)\kappa\lambda/\varrho$ behaves  the role of a cosmological constant.
The behaviors of metric function $A(\rho)$ listed in Tab. II. It shows that all solutions are asymptotic to dS/AdS space when $\Lambda>0$ or $\Lambda<0$, and constant $\rho_1\geq0$ corresponding to wormhole and $\rho_1<0$ to black hole solution.
 \begin{table}
 \caption{Asymptotic behaviors of metric function $A(\rho)$ in Einstein-bumblebee gravity coupled with a phantom field under the linear bumblebee potential $V=\lambda x/2$.}\label{tab2}
\begin{center}
\begin{tabular}{|c|c|c|c|}
\hline  $\rho_1$ & large $\rho$ & little $\rho$ & kinds of solution  \\
\hline $\rho_1=0$ & $\rho\rightarrow\pm\infty, A(\rho)\rightarrow-\Lambda\rho^2$ & $\rho\rightarrow\pm0, A(\rho)\rightarrow(1-p^2\Lambda)/3$&regular AdS wormhole\\
$\rho_1>0$ & $\rho\rightarrow\pm\infty, A(\rho)\rightarrow-\Lambda\rho^2$ & $\rho\rightarrow\pm0, A(\rho)\rightarrow\pm\infty$&Schwarzschild-AdS-like wormhole\\
$\rho_1<0$ & $\rho\rightarrow\pm\infty, A(\rho)\rightarrow-\Lambda\rho^2$ & $\rho\rightarrow\pm0, A(\rho)\rightarrow\mp\infty$& BP-dS/AdS black hole\\
\hline
\end{tabular}
\end{center}
\end{table}
The phantom field and its potential are,
\begin{eqnarray}\label{}
\Phi'^2=2(1+\ell)\Big[\frac{p^2}{r^4}+b_0^2\frac{\kappa\lambda}{ A(\rho)}\Big],\qquad \mathcal{V}=\frac{2p^2}{r^4}A(\rho)+\Lambda,
\end{eqnarray}
which are both asymptotical to a constant, i.e., $\Phi\rightarrow$const. and $\mathcal{V}\rightarrow$ const.  at $\rho\rightarrow\pm\infty$.
One can see that in this case, the contents of the phantom field depend on the parameter $p$ and Lagrange-multiplier $\lambda$.
\subsubsection{regular AdS wormhole solution}
If $\rho_1=0$, then $A(\rho)$ becomes
\begin{eqnarray}\label{}
A(\rho)=\big(1+\frac{p^2}{\rho^2}\big)\Big(1-\frac{p}{\rho}
\arctan\frac{\rho}{p}-\frac{\Lambda}{3}\rho^2\big).
\end{eqnarray}
If $\Lambda<0$, then $(1-p^2\Lambda)/3<A(\rho)<+\infty$, it is a wormhole solution with the throat $r=p$, which connects two AdS universe.  Its Ricci scalar $R$ is
\begin{eqnarray}\label{}
R=\frac{2}{(1+\ell)(\rho^2+p^2)}\Big[\ell+\Lambda(p^2+2\rho^2)-\frac{p^2}{\rho^2}\big(1+\frac{6p^2}{\rho^2}\big)
+\frac{3p^3}{\rho^3}\big(1+\frac{2p^2}{\rho^2}\big)\arctan\frac{\rho}{p}
\Big].
\end{eqnarray}
At the origin $\rho\rightarrow0$, it becomes,
\begin{eqnarray}\label{}
R=\frac{2}{(1+\ell)p^2}\big(\frac{6}{5}+\ell+\Lambda p^2\big),
\end{eqnarray}
which is finite. Then this AdS wormhole is regular everywhere and has no singularity.
\subsubsection{Bumblebee-phantom(BP)-dS/AdS black hole solution}
If $\rho_1=\pi p/2-2M<0$, then $A(\rho)$ becomes,
\begin{eqnarray}\label{metricf2}
A(\rho)=\big(1+\frac{p^2}{\rho^2}\big)\Big[1-\frac{2M}{\rho}-\frac{ \Lambda}{3}\rho^2+\frac{p}{\rho}\big(\frac{\pi}{2}
-\arctan\frac{\rho}{p}\big)\Big],
\end{eqnarray}
which is a  bumblebee-phantom(BP)-dS/AdS  black hole solution.
When  $\rho\rightarrow\infty$, the metric function (\ref{metricf2}) is asymptotical to the Schwarzschild (anti-)de Sitter (dS/AdS) black hole solution $A(\rho)\rightarrow1-2M/\rho-\Lambda\rho^2/3$ and, the phantom field $\Phi\rightarrow$ constant, and its potential  $\mathcal{V}\rightarrow\Lambda$ constant also. It is easy to see that the phantom potential $\mathcal{V}$ behaves as a cosmological constant to this BP-dS/AdS black hole.

From the metric (\ref{metric1}), one can see that the bumblebee Schwarzschild (anti-)de Sitter(dS/AdS)-like black solution \cite{ding2023,maluf} can be recovered when $p\rightarrow0$, \begin{eqnarray}\label{}A(\rho)=1-2M/\rho-(1+\ell)\frac{\Lambda_e}{3}\rho^2,\end{eqnarray} with effective cosmological constant $\Lambda_e=\kappa \lambda/\varrho$. And the phantom potential becomes,
\begin{eqnarray}\label{}
\mathcal{V}=(1+\ell)\frac{\kappa\lambda}{\varrho}=(1+\ell)\Lambda_e,
\end{eqnarray}
which is the cosmological constant. However, in this case, the phantom field $\Phi$,
 \begin{eqnarray}\label{}
\Phi'^2=2(1+\ell)b_0^2\frac{\kappa\lambda}{A(\rho)},
\end{eqnarray}which is not a constant. This phantom field is related to Lagrange-multiplier $\lambda$ field which is an auxiliary field \cite{ding2023}.

\section{Smarr formula and first law for BP black hole}
In this section, we derive the Smarr formula by introducing Killing potential $\omega^{ab}$ and construct the first law of black hole thermodynamics for BP black hole. Next, we compare them to those of no Lorentz violation case---the regular phantom black hole reported in Ref. \cite{bronnikov}, and find the effect of Lorentz violation.

Suppose that $\mathcal{M}$ is this $4$-dimensional spacetime satisfying the Einstein equations, $\xi^a=(1,0,0,0)$ is a Killing vector on $\mathcal{M}$, timelike near infinity. In $\mathcal{M}$,  there is a spacelike hypersurface $\tilde{S}$ with a co-dimension 2-surface boundary $\partial \tilde{S}$, and $\xi^a$ is normal to the $\tilde{S}$. The boundary $\partial \tilde{S}$ has two components: an inner boundary at the event horizon $\partial \tilde{S}_h$ and an outer boundary at infinity $\partial \tilde{S}_\infty$.
We can integrate the Killing equation $\nabla_b(\nabla^b\xi^a)=-R^a_b\xi^b$ over this hypersurface $\tilde{S}$,
\begin{eqnarray}\label{intone}
\int_{\partial \tilde{S}}\nabla^b\xi^ad\Sigma_{ab}=-\int_{\tilde{S}}R^a_b\xi^bd\Sigma_a,
\end{eqnarray}
where $d\Sigma_{ab}$ and $d\Sigma_{a}$ are the surface elements of $\partial \tilde{S}$ and $\tilde{S}$, respectively.
The non-vanishing components of the tensor $\nabla^b\xi^a$ and $R_t^t$ are given by
\begin{eqnarray}\label{Rtt}
\nabla^\rho\xi^t=-\nabla^t\xi^\rho=-\frac{A'(\rho)}{2(1+\ell)},\quad R_t^t=\frac{2p^2A(\rho)}{(1+\ell)r^4}.
\end{eqnarray}
Since Ricci tensor $R^a_b$ is nonzero, Gauss's law cannot be used to the right side of Eq. (\ref{intone}). However in Ref.  \cite{kastor}, Komar integral relation still holds with $R_{t}^t=\Lambda \neq0$, i.e., dS/AdS black hole spacetime by introducing an anti-symmetric Killing potential $\omega^{ab}$ which can be obtained according to relation $\xi^b=\nabla_a\omega^{ab}$.
In Ref. \cite{ding2023}, we used it to study thermodynamics for high dimensional dS/AdS bumblebee black hole.  For the present static Killing vector $\xi^a$, we have nonzero components of $\omega^{ab}$ that,
\begin{eqnarray}\label{}
\omega^{\rho t}=-\omega^{t\rho}=\frac{\rho(\rho^2+3p^2)}{3(\rho^2+p^2)}+ \frac{\alpha}{\rho^2+p^2},
\end{eqnarray}
where $\alpha$ is an integral constant. However, $R_t^t$ in Eq. (\ref{Rtt}) is not a constant but a function of $\rho$. Therefore, one should
introduce another function $g(\rho)$ to lead to the relation that,
\begin{eqnarray}\label{}
\nabla_b[g(\rho)\omega^{ab}]=R^t_t\nabla_b\omega^{ab}.
\end{eqnarray}
We find that,
\begin{eqnarray}\label{}
g(\rho)=\frac{3}{(1+\ell)(3p^2+\rho^2)}\Big[\frac{p}{\rho}\arctan\frac{\rho}{p}
-\frac{p^2}{\rho^2+p^2}A(\rho)\Big].
\end{eqnarray}
Lastly, the Eq. (\ref{intone}) can be rewritten as
\begin{eqnarray}\label{inttwo}
\frac{1}{4\pi}\int_{\partial \tilde{S}}\left(\nabla^b\xi^a+g(\rho)\omega^{ab}\right)d\Sigma_{ab}=0,
\end{eqnarray}
which is multiplied by the normalization factor $1/4\pi$ and called the Komar integral relation.

The closed 2-surface $\partial \tilde{S}$ has two parts, horizon $\partial \tilde{S}_h$ and infinite $\partial \tilde{S}_\infty$,
so Eq. (\ref{inttwo}) can be rewritten as
\begin{eqnarray}\label{intthree}
\frac{1}{4\pi}\int_{\partial \tilde{S}_\infty}\left[\nabla^b\xi^a+g(\rho)\omega^{ab}\right]d\Sigma_{ab}=\frac{1}{4\pi}\int_{\partial \tilde{S}_h}\left[\nabla^b\xi^a+g(\rho)\omega^{ab}\right]d\Sigma_{ab}.
\end{eqnarray}
If we use the 2-surface element $d\Sigma_{\rho t}=-\sqrt{1+\ell}(\rho^2+p^2)d\Omega/2$, which is slightly modified by the factor $\sqrt{1+\ell}$, the left and right hand sides of this integral are,
 \begin{eqnarray}\label{}
&&\frac{1}{4\pi}\int_{\partial \tilde{S}_\infty}\left[\nabla^b\xi^a+g(\rho)\omega^{ab}\right]d\Sigma_{ab}
=\frac{1}{\sqrt{1+\ell}}\big(M-\frac{\pi}{2}p\big),\\&& \frac{1}{4\pi}\int_{\partial \tilde{S}_h}\left[\nabla^b\xi^a+g(\rho)\omega^{ab}\right]d\Sigma_{ab}
=\frac{1}{\sqrt{1+\ell}}\big(2TS-p\arctan\frac{\rho_h}{p}\big).
\end{eqnarray}
So the integral (\ref{intthree}) can give,
\begin{eqnarray}\label{smar}
M=2TS+p\big(\frac{\pi}{2}-\arctan\frac{\rho_h}{p}\big),
\end{eqnarray}
where the entropy $S$ is a quarter of horizon area $\tilde{A}$ \cite{rodrigues},
\begin{eqnarray}\label{}
\tilde{A}=\int_{horizon}\sqrt{g_{\theta\theta}g_{\varphi\varphi}}d\theta d\varphi=4\pi\sqrt{1+\ell}(\rho_h^2+p^2).
\end{eqnarray}
The phantom charge \footnote{\textcolor[rgb]{0.00,0.00,1.00}{From the Eq. (\ref{metricf}), the parameter $p$ is an integral constant and $p<4M/\pi$, which is like the role of the electric charge $Q$ in Reissner-Nordst\"{o}m black hole.} } \cite{gao} is,
\begin{eqnarray}\label{}
Q_p=\frac{1}{4\pi}\int d^2\Sigma^\rho\nabla_\rho\Phi=\sqrt{2(1+\ell)}p,
\end{eqnarray}
and defining phantom effective potential,
\begin{eqnarray}\label{}
V_p=\frac{1}{2\sqrt{2(1+\ell)}}\big(\frac{\pi}{2}-\arctan\frac{\rho_h}{p}\big),
\end{eqnarray}
then Eq. (\ref{smar}) can be rewritten as the Smarr formula,
\begin{eqnarray}\label{}
M=2TS+2V_pQ_p.
\end{eqnarray}

The first law of black hole thermodynamics can be constructed by the following method as,
\begin{eqnarray}\label{}
dM=\left(\frac{\partial M}{\partial S}\right)_{Q_p}dS+\left(\frac{\partial M}{\partial Q_p}\right)_SdQ_p.
\end{eqnarray}
From the black hole's horizon Eq. (\ref{horizon}), one can write the black hole mass as following,
\begin{eqnarray}\label{}
M=\frac{\rho_h}{2}+\frac{p}{2}\big(\frac{\pi}{2}-\arctan\frac{\rho_h}{p}\big)
=\frac{1}{2}\sqrt{\frac{S}{\pi\sqrt{1+\ell}}-\frac{Q_p^2}{2(1+\ell)}}+V_pQ_p.
\end{eqnarray}
Lastly, the first law is,
\begin{eqnarray}\label{}
dM=TdS+V_pdQ_p.
\end{eqnarray}
One can see that the conventional first law still holds for this Lorentz violation BP black hole. However, for the case of no Lorentz violation---the regular phantom black hole in Ref. \cite{bronnikov}, the Smarr formula is $M=2TS+(2M-\rho_h)$ which is very strange to us, and the first law cannot be constructed (see the appendix C for detail). So one can conclude that, the appearance of Lorentz invariance violation improves the structure of the phantom black hole.

\begin{figure}[ht]
\begin{center}
\includegraphics[width=7.0cm]{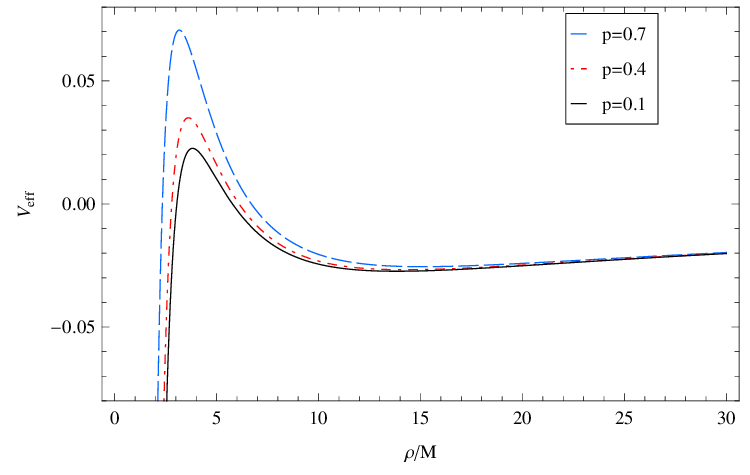}
\caption{Evolution of the gravitational potential for the equatorial circular orbit with $\ell=0.2, L/M=4.2$.}\label{fp1}
 \end{center}
 \end{figure}

\textcolor[rgb]{0.00,0.00,1.00}{\section{Existence of a stable circular orbits for BP black hole}}
In this section we study some effects might manifest in astrophysical observations for BP black hole.
The affinely parameterized geodesic equation for timelike particles in the equatorial plane $\theta=\pi/2$ is described by the Lagrangin \cite{james}
\begin{eqnarray}\label{}
\mathcal{L}=-\frac{A(\rho)}{2}\left(\frac{dt}{ds}\right)^2+\frac{1+\ell}{2A(\rho)}
\left(\frac{d\rho}{ds}\right)^2+\frac{\rho^2+p^2}{2}\left(\frac{d\varphi}{ds}\right)^2
=-\frac{1}{2},
\end{eqnarray}
where,
\begin{eqnarray}\label{}
A(\rho)=\big(1+\frac{p^2}{\rho^2}\big)\big(1-\frac{H}{\rho}\big),\qquad H=2M-p\big(\frac{\pi}{2}-\arctan \frac{\rho}{p}\big).
\end{eqnarray}
There are both conservation quantities: particles' energy $E$  and angular momentum $L$ per unit rest mass,
\begin{eqnarray}\label{}
A(\rho)\left(\frac{dt}{ds}\right)=E,\qquad (\rho^2+p^2)\left(\frac{d\varphi}{ds}\right)=L.
\end{eqnarray}
By defining constant $\mathcal{E}=(E^2-1)/2(1+\ell)$, we have,
\begin{eqnarray}\label{}
\mathcal{E}=\frac{1}{2}\left(\frac{d\rho}{ds}\right)^2+V_{eff},
\end{eqnarray}
where the effective gravitational potential is,
\begin{eqnarray}\label{}
V_{eff}=\frac{1}{2(1+\ell)}\Big[-\frac{H}{\rho}+\big(1-\frac{H}{\rho}\big)\frac{p^2+L^2}{\rho^2}\Big],
\end{eqnarray}
which is displayed in Fig. \ref{fp1}.
We can see that for large values of $\rho$, this potential is close to the Newtonian effective potential $-M/\rho$ modified by the LV constant $\ell$;  and at the horizon $\rho_h$, it is a finite value, i.e.,
\begin{eqnarray}\label{}
\rho\rightarrow\infty, V_{eff}\rightarrow-\frac{M}{(1+\ell)\rho};\qquad\rho=\rho_h, V_{eff}=-\frac{1}{2(1+\ell)},
\end{eqnarray}
which show the effect of LV constant $\ell$, and are independent on the phantom charge $Q_p$.
Fig. \ref{fp1} shows that there are both stationary points: the radii $\rho_{min}$ for the local minimum of the potential $V_{eff}$ and $\rho_{max}$ for the maximum one which determined by the equation $dV_{eff}/d\rho=0$. Point $\rho_{max}$ is unstable, while the point $\rho_{min}$ is an attractor, around which there exist periodic solutions $\rho(s)$, i.e., the existence of stable circular orbits. 
Fig. \ref{fp1} also shows that, for a given particle, i.e., $L$ is given, the radii $\rho_{min}$ increases with the phantom parameter $p$.

$\rho_{min}$ decreases with decreasing particles' angular momentum $L/M$, but it is impossible at arbitrary small value. If $L<L_{min}$, an inwardly directed particle falls all the way to the origin of the black hole. There exists the innermost stable circular orbit(ISCO) $\rho_{isco}$ determined by the equation $d^2V_{eff}/d\rho^2=0$ and displayed in Tab. III.
 \begin{table}
 \caption{Some values of horizon radii  $\rho_h$, ISCO radii $\rho_{isco}$ and particles' minimal angular momentum $L_{min}$ for BP black hole with different $p$. They all aren't dependent on LV constant $\ell$. Note that $p<4M/\pi$.}\label{}
\begin{center}
\begin{tabular}{|c|c|c|c|c|c|c|}
\hline  $p/M$ &0.05 &0.1&0.4&0.7 & 1&1.2 \\
\hline $\rho_h/M$ &1.999 &1.995&1.918&1.731 &1.369&0.864\\
 $\rho_{isco}/M$ &5.996 &5.985&5.754&5.204 &4.170&2.798\\
  $L_{min}/M$ &3.463 &3.458&3.361&3.128 &2.678&2.057\\
 \hline 
\end{tabular}
\end{center}
\end{table}
Table III shows that the horizon $\rho_h$,  ISCO radii $\rho_{isco}$ and the smallest particles' angular momentum $L_{min}$ decrease with increasing phantom parameter $p$. When $p\rightarrow0$,   $\rho_h\rightarrow2M$, $\rho_{isco}\rightarrow6M$ and $L_{min}\rightarrow\sqrt{12}M$.
 
In a conclusion, from the view of astrophysical observations, the stable circular orbit still exists, the effects of the Lorentz violation and the phantom field on the gravitational potential and particle motion are obvious and observable.
For the connection to  constraints from gravitational wave data or cosmological observations, is left to study in the near future.

\section{Summary}

In this paper, we have  studied Einstein-bumblebee gravity theory minimally coupling to an external  matter---conventional non-phantom scalar or exotic phantom scalar field. In Einstein-bumblebee grvity theory, an constant vector $b^\mu$ ($b^\mu b_\mu=$ constant) is introduced to violate the Lorentz symmetry spontaneously. This constant vector  $b^\mu$ is the nonzero vacuum expectation of a dynamic vector field $B^\mu$---the bumblebee field. We find that the bumblebee field can influence the gravitational field solutions via the coupling constant $\varrho$ ($\varrho$ is nonzero and $\varrho\neq-1/b_0^2$) and the specific forms of bumblebee potential $V(x)$.  And the external scalar field $\Phi$ can also affect the gravitational field solutions(give them a hair), but its contents and the forms of its potential $\mathcal{V}(\Phi)$ are determined by the gravitational field. Besides, this scalar field can sustain the exist of a wormhole, its potential can be use of a cosmological constant.

The first Lorentz invariance violation effect is that, it can change the so called no-hair theorem: when the bumblebee coupling constant $\ell>-1$, the phantom field with negative kinetic energy is admissible but the conventional positive kinetic energy field is forbidden; when the coupling constant $\ell<-1$, the phantom field with negative kinetic energy is forbidden but the conventional positive kinetic energy field is admissible.

Then we have derived some wormhole and black hole solutions under the two forms of the bumblebee potential. When under the quadratic form of the bumblebee potential $V=kx^2/2$, the metric function $A(\rho)$ of the solutions are all asymptotical to unit $A(\rho)\rightarrow1$,  and three specific examples are given: i) Ellis-bumblebee-phantom wormhole solution which is regular everywhere and has no singularity; ii) Schwarzschild-like wormhole solution which has a naked singularity at the origin; iii)  bumblebee-phantom(BP) black hole solution which has a horizon located at $\rho_h=2M-p(\pi/2-\arctan \rho_h/p)$. For this hairy BP black hole, the form of phantom charge $Q_p=\sqrt{2(1+\ell)}p$, corresponding potential $V_p=(\pi/2-\arctan \rho_h/p)/2\sqrt{2(1+\ell)}$. The Lorentz invariance violation effect is that, it can improve the structure of this phantom hairy black hole---the conventional Smarr formula and the first law of black hole thermodynamics still hold. On the contrary,  for the case of no Lorentz invariance violation, i.e., the regular phantom black hole reported in Ref. \cite{bronnikov}, the first law cannot be constructed at all.

When under the linear Lagrange multiplier potential $V=\lambda x/2$, the solutions are all asymptotically to dS/AdS space and a specific example are given: i) regular AdS wormhole solution; ii) bumblebee-phantom-dS/AdS black hole solution which can recover the bumblebee Schwarzschild dS/AdS-like black hole solution.

\begin{acknowledgments}  This work was supported by the Scientific Research Fund of the Hunan Provincial Education Department under No. 19A257 and No. 19A260, the National Natural Science Foundation (NNSFC)
of China (grant No. 11247013), Hunan Provincial Natural Science Foundation of China grant No. 2015JJ2085.
\end{acknowledgments}

\appendix\section{Some contraction conventions and some nonzero components of the given tensors}
In this appendix, we show some contraction conventions with a given metric and some the nonezero components of Einstein's tensor and the energy momentum tensor of the bumblebee field with the metric (\ref{metric}).

For the sign of a metric $g_{\mu\nu}$, there are two kinds: $(+---)$ and $(-+++)$; for the Ricci tensor $R_{\mu\nu}$, there are also two kinds of contraction:  contraction of the first and third or, the first and fourth index of the Riemann tensor, i.e., $R_{\mu\nu}=R^\sigma_{\;\mu\sigma\nu}$ or $R_{\mu\nu}=R^\sigma_{\;\mu\nu\sigma}$. Different conventions will give different sign of a resulting tensor or Ricci scalar. We list some signs as following Tab. IV.
 \begin{table}
 \caption{Some contraction conventions with a given metric.}\label{}
\begin{center}
\begin{tabular}{|c|c|c|c|}
\hline  $(-+++),R^\sigma_{\;\mu\sigma\nu}$ &$(-+++),R^\sigma_{\;\mu\nu\sigma}$ & $(+---),R^\sigma_{\;\mu\nu\sigma}$ & $(+---),R^\sigma_{\;\mu\sigma\nu}$  \\
\hline $g_{\mu\nu}$ & $g_{\mu\nu}$ & $-g_{\mu\nu}$& $-g_{\mu\nu}$\\
$R_{\mu\nu}$ & $-R_{\mu\nu}$ & $-R_{\mu\nu}$& $R_{\mu\nu}$\\
$R$ & $-R$ & $ R$& $-R$\\
$G_{\mu\nu}$ & $-G_{\mu\nu}$ & $ -G_{\mu\nu}$& $G_{\mu\nu}$\\
\hline
\end{tabular}
\end{center}
\end{table}

 The nonezero components of Einstein's tensor and the energy momentum tensor of the bumblebee field with the metric (\ref{metric}) and with the contraction of $R_{\mu\nu}=R^\sigma_{\;\mu\nu\sigma}$ are as following:
\begin{eqnarray}
&&G_{00}=\frac{e^{2\phi-2\psi}}{r^2}\Big[-e^{2\psi}+r'^2-2rr'\psi'+2rr''\Big],\\
&&G_{11}=\frac{1}{r^2}\Big[e^{2\psi}-r'2+2rr'\phi'\Big],\\
&&G_{22}=-e^{-2\psi}\Big[rr'(\phi'-\psi')
+r^2(\phi''+\phi'^2-\phi'\psi')+rr''\Big],\\
&&R_{11}=-\frac{2r'}{r}\psi'+2\frac{r''}{r}+(\phi''+\phi'^2-\phi'\psi').
\end{eqnarray}
$C_{\mu\nu}$  are
\begin{eqnarray}
&&C_{00}=\frac{\varrho b_0^2e^{2\phi-2\psi}}{2}R_{11},\quad
C_{11}=2\kappa V'b_0^2e^{2\psi}+\frac{3\varrho b_0^2}{2}R_{11},\quad
C_{22}=-\frac{\varrho b_0^2}{2}r^2e^{-2\psi}R_{11}.
\end{eqnarray}
$\bar B_{\mu\nu}$  are
\begin{eqnarray}
&&\bar B_{00}=-\frac{\varrho b_0^2e^{2\phi-2\psi}}{2r^2}\Big[2r'^2-2rr'\psi'-r^2(\phi''+\phi'^2
-\phi'\psi')\Big],\\
&&\bar B_{11}=\frac{\varrho b_0^2}{2r^2}\Big[2r'^2-2rr'(\psi'-2\phi')++2rr''+r^2(\phi''+\phi'^2-\phi'\psi')\Big],\\
&&\bar B_{22}=\frac{\varrho b_0^2e^{-2\psi}}{2}\Big[2rr'\phi'+r^2(\phi''
+\phi'^2-\phi'\psi')\Big].
\end{eqnarray}

\section{An examples for Ref. \cite{ovgun}}
In Ref. \cite{ovgun}, \"{O}vg\"{u}n {\it et al} studied the traversable wormholes in bumblebee gravity with the external isotropic matter $(-\tilde{\rho},\tilde{p},\tilde{p},\tilde{p})$, where $\tilde{\rho}$ is the energy density of the external matter, and $\tilde{p}$ is the pressure  of the matter. They used the metric ansatz,
\begin{eqnarray}
ds^2=-dt^2+\frac{dr^2}{1-\frac{W(r)}{r}}+r^2d\Omega^2,
\end{eqnarray}
and found that\begin{eqnarray}
W(r)=\frac{1}{1+\ell}\left[\ell r+r_0\big(\frac{r_0}{r}\big)^{-\frac{5\ell+3}{3\ell+1}}\right],
\end{eqnarray}
where $r_0$ is a constant and stands for throat radius. This metric can be rewritten as
\begin{eqnarray}
ds^2=-dt^2+\frac{(1+\ell)dr^2}{f(r)}+r^2d\Omega^2,\qquad f(r)=\left[1-\big(\frac{r_0}{r}\big)^{-\frac{2\ell+2}{3\ell+1}}\right].
\end{eqnarray}
When the coupling constant $\ell=-2$, the shape function $f(r)$ and the energy density are \begin{eqnarray}
f(r)=1-\big(\frac{r}{r_0}\big)^{\frac{2}{5}},\qquad \tilde{\rho}=\frac{7}{5\kappa r_0^2}(r_0r^2)^{-\frac{4}{5}}>0.
\end{eqnarray}
So this matter is conventional positive energy matter.

\section{Komar integral for phantom regular black hole in Ref. \cite{bronnikov}}
In Ref. \cite{bronnikov}, Bronnikov and Fabris reported a regular phantom black hole which was reconsidered by Ding {\it et al} in Ref. \cite{ding2013}. Its metric is,
\begin{eqnarray}\label{}
&&ds^2=f(\rho)dt^2-\frac{1}{f(\rho)}d\rho^2-(\rho^2+p^2)d\Omega^2,
\end{eqnarray}
where the function $f(\rho)$ is,
\begin{eqnarray}\label{}
f(\rho)=1-\frac{3M}{p}\Big[\big(1+\frac{\rho^2}{p^2}\big)\big(\frac{\pi}{2}
-\arctan\frac{\rho}{p}\big)-\frac{\rho}{p}\Big].
\end{eqnarray}
The black hole horizon $\rho_h$ is the biggest root of the equation that,
\begin{eqnarray}
\frac{3M}{p}\Big[\big(1+\frac{\rho_h^2}{p^2}\big)\big(\frac{\pi}{2}
-\arctan\frac{\rho_h}{p}\big)-\frac{\rho_h}{p}\Big]=1.\end{eqnarray}
Its Hawking temperature $T$ is,
\begin{eqnarray}
T=\frac{3M-\rho_h}{2\pi (p^2+\rho_h^2)}.\end{eqnarray}
The non-vanishing components of the tensor $\nabla^b\xi^a$ and $R_t^t$ are given by
\begin{eqnarray}\label{}
\nabla^\rho\xi^t=-\nabla^t\xi^\rho=\frac{f'(\rho)}{2},\quad R_t^t=\frac{3M}{pr^2}\Big[\big(1+\frac{3\rho^2}{p^2}\big)\big(\frac{\pi}{2}
-\arctan\frac{\rho}{p}\big)-\frac{3\rho}{p}\Big].
\end{eqnarray}
The function $g(\rho)$ is
\begin{eqnarray}\label{}
g(\rho)=\frac{3}{(3p^2+\rho^2)}(1-f).
\end{eqnarray}
One uses the 2-surface element $d\Sigma_{\rho t}=-(\rho^2+p^2)d\Omega/2$, the left and right hand sides of this integral are,
 \begin{eqnarray}\label{}
&&\frac{1}{4\pi}\int_{\partial \tilde{S}_\infty}\left[\nabla^b\xi^a+g(\rho)\omega^{ab}\right]d\Sigma_{ab}
=\big(M-2M\big),\\&& \frac{1}{4\pi}\int_{\partial \tilde{S}_h}\left[\nabla^b\xi^a+g(\rho)\omega^{ab}\right]d\Sigma_{ab}
=\big(2TS-\rho_h\big).
\end{eqnarray}
So the integral (\ref{intthree}) can give,
\begin{eqnarray}\label{}
M=2TS+\big(2M-\rho_h\big),
\end{eqnarray}
which is very strange to us. And we find that the first law cannot be constructed.

\end{document}